\documentclass[manuscript,acmsmall,nonacm]{acmart}

\setcopyright{none}

\AtBeginDocument{%
  }

\setcopyright{acmlicensed}
\copyrightyear{2018}
\acmYear{2018}
\acmDOI{XXXXXXX.XXXXXXX}
\acmConference[Conference acronym 'XX]{Make sure to enter the correct
  conference title from your rights confirmation email}{June 03--05,
  2018}{Woodstock, NY}
\acmISBN{978-1-4503-XXXX-X/2018/06}

\usepackage{multirow} 
\usepackage{tcolorbox} 
\usepackage{lipsum} 

\usepackage{multirow}
\usepackage{graphicx}
\usepackage{tabularx,tabulary}
\usepackage{xcolor}

\usepackage[normalem]{ulem}
\useunder{\uline}{\ul}{}
\usepackage{array}
\usepackage{pdflscape}  
\usepackage[table]{xcolor}  

\begin{document}


\title{Children's Privacy Reasoning About Deceptive Patterns in Mobile Apps and the Role of Parental Mediation}

\author{Justin Eric Chen}
\email{jechen4@illinois.edu}
\affiliation{%
  \department{School of Information Sciences}
  \institution{University of Illinois Urbana-Champaign}
  \city{Champaign}
  \state{Illinois}
  \country{USA}}

\author{Ece Gumusel}
\email{ece.g@rutgers.edu}
\affiliation{%
  \department{Department of Library and Information Science, School of Communication and Information}
  \institution{Rutgers, The State University of New Jersey}
  \city{New Brunswick}
  \state{New Jersey}
  \country{USA}}

\author{Cora Yilin Zhu}
\email{corayilinzhu@gmail.com}
\affiliation{%
  \department{Entertainment Technology Center}
  \institution{Carnegie Mellon University}
  \city{Pittsburgh}
  \state{Pennsylvania}
  \country{USA}}

\author{Diana Ngo}
\email{dngo25@illinois.edu}
\affiliation{%
  \department{School of Information Sciences}
  \institution{University of Illinois Urbana-Champaign}
  \city{Champaign}
  \state{Illinois}
  \country{USA}}

\author{Sang-Hwa Oh}
\email{sanghwa2@illinois.edu}
\affiliation{%
  \department{College of Media}
  \institution{University of Illinois Urbana-Champaign}
  \city{Champaign}
  \state{Illinois}
  \country{USA}}

\author{Kyrie Zhou}
\email{kyrie.zhou@utsa.edu}
\affiliation{%
  \department{Department of Information Systems and Cybersecurity}
  \institution{University of Texas at San Antonio}
  \city{San Antonio}
  \state{Texas}
  \country{USA}}

\renewcommand{\shortauthors}{Chen et al.}


\begin{abstract}
Children increasingly encounter online interfaces that use deceptive design to influence decisions about privacy, permissions, purchases, and content sharing. While HCI research has examined deceptive patterns among adults, less is known about how children recognize and respond to them, or how their families' privacy practices relate to these responses. We conducted a multi-method qualitative study with $N=18$ participants ($9$ child--guardian dyads; children aged 8--12), combining interviews, scenario-based walkthroughs with think-aloud, video analysis, and questionnaires. Children completed six interface scenarios featuring deceptive patterns, including obstruction, nagging, sneaking (preselection), interface interference, and forced action. Children's recognition varied widely. In several families, it appeared to relate to what guardians' privacy conversations addressed. Children whose guardians explained how apps request and use data often reasoned about the designs, while those given mainly rules about what to avoid tended to miss preselection and visual steering, though this pattern was not uniform. We discuss how families may move from privacy rules to privacy reasoning, how mediation can balance protection with children's agency, and how this responsibility can be shared beyond parents.
\end{abstract}



\maketitle




\section{Introduction}
\label{introduction}

The preteen years, roughly ages 9 to 12, are an important period of  development, as children begin to think more abstractly and to reason  about hypothetical situations \cite{jordan2017growing}. During this stage, children also begin to develop a sense of identity separate from their  parents and siblings, along with their own personality, beliefs, and moral  compass \cite{moreno2012social}. As preteens begin to distance themselves  from their parents, this transition can lead to both conflict and greater self-efficacy \cite{lerner2009handbook}. Online spaces can support this  process, giving children an opportunity to explore different aspects of their identity \cite{weinstein2014personal}. However, as children engage with apps and social media, they are also exposed to privacy risks and 
manipulative design practices that shape how they make decisions and  interact with digital platforms \cite{radesky2022prevalence,zhao2020data}. Thus, parents and guardians face challenges in managing their children's  app and social media use, as they must balance allowing children to  express themselves and exercise autonomy with keeping them safe in  online spaces.

Parents may attempt to protect their children by restricting, regulating, or discussing their mobile application use \cite{beyens2022social,george2015seven}. However, many of these applications rely on deceptive patterns, interface designs that manipulate or deceive users into taking actions they might not otherwise choose \cite{mathur2019dark,gray2018dark,schaffner2022understanding,di2020ui,sin2025dark,Brignull_2023}. Because these patterns exploit fast, automatic thinking rather than deliberate reasoning \cite{Bsch2016TalesFT,tversky1974judgment}, they are subtle and hard to detect \cite{gray2018dark}. This makes it harder for parents to protect their children online.

Because of the risks posed by deceptive patterns, a substantial body of research has examined this domain, including their prevalence and characteristics \cite{mathur2019dark,mildner_2023,Mildner_2023_gov,essmeyer2026bridge}, as well as legislative efforts to regulate them \cite{yi2025mapping,chen2026regulating}. More recently, deceptive pattern research has expanded to examine how individual characteristics and vulnerability may shape users' ability to recognize and respond to these deceptive practices. For example, Mildner et al. \cite{mildner2025comparative} investigated how ADHD may affect an individual's ability to recognize deceptive patterns. Similarly, recent studies have examined children's ability to recognize and respond to deceptive patterns, as well as how children interact with these manipulative interface designs \cite{sousa2023dark,renaud2024we,schafer2024growing,schafer2025if,arif2026understanding}. For example, Arif et al. \cite{arif2026understanding} found that children aged 11–13 can recognize when video game designs are manipulative, and Sch{\"a}fer et al. \cite{schafer2025if} found that both children (10–12) and adolescents (16–18) understand that interface designs can influence people's decisions, but they demonstrate that understanding in different ways when asked to create deceptive  designs. Similarly, Renaud et al. \cite{renaud2024we} found that 11–12-year-old children were highly aware of online deception and generally perceived deceptive actors as having malicious intentions. 

These studies consistently find that children can recognize deceptive patterns, but offer limited insight into why some children recognize them while others do not. One possible explanation lies in the parental mediation children experience. Restrictive mediation may reduce children's exposure to online risks but limit their opportunities to develop the skills to recognize and respond to such risks independently \cite{livingstone2017maximizing,rodriguez2018study}, whereas active mediation, in which parents discuss media content with children and build their critical viewing skills \cite{rasmussen2013theoretical}, may better support children in evaluating persuasive content on their own \cite{buijzen2005parental}. Yet mediation research has focused on screen time, content exposure, and advertising, and little is known about how the guidance children receive at home relates to how they respond to deceptive patterns. This gap matters because preteens, as they gain autonomy and lose close supervision, increasingly face these designs alone and must rely on what they have learned rather than on a parent's intervention in the moment. We therefore conducted an exploratory study of how children respond to privacy-related deceptive patterns and how their responses relate to the guidance they receive at home. We address the following research questions:

\begin{enumerate}
    \item[\textbf{RQ1}] How do children aged 8–12 notice, interpret and respond to privacy-related deceptive patterns in a mobile app? 
    \item[\textbf{RQ2}] How do parents’ privacy views and mediation practices relate to how their children recognize and respond to deceptive patterns?
\end{enumerate}

To answer these questions, we conducted a multi-method study with nine 
parent--child dyads. Children completed an interactive walkthrough of six 
privacy-related deceptive patterns while thinking aloud, along with 
interviews before and after. Parents were interviewed separately about 
how they guide their children's app use.

\section{Related Work}

\subsection{Deceptive Patterns and Children's Susceptibility}

Dark patterns are user interfaces intentionally designed to confuse, mislead, or manipulate users into taking actions they might not otherwise choose \cite{luguri2021shining,Brignull_2023,Mathur_2021}. We use the term \emph{deceptive patterns}, following Brignull's revised terminology \cite{Brignull_2023}, which has been increasingly adopted in HCI research. These designs are particularly concerning in privacy contexts because they can obscure how personal information is collected and used, leading users to disclose more information than intended \cite{waldman2020cognitive,Bsch2016TalesFT}.

Deceptive patterns often exploit cognitive biases and limitations in human decision-making. Kahneman's \cite{kahneman2011thinking} dual-process theory distinguishes between System 1, which involves fast, automatic thinking, and System 2, which involves slower, deliberate reasoning. Deceptive patterns can exploit reliance on System 1 by encouraging quick decisions without careful consideration of their consequences \cite{Bsch2016TalesFT,tversky1974judgment}. This is particularly concerning for children, whose inhibitory control, or ability to suppress automatic responses in favor of deliberate ones, is still developing \cite{davis2011exercise,Macdonald03092014,Sadeghi25012022}. Consequently, children may struggle to interrupt automatic responses and consider the privacy implications of manipulative interfaces \cite{oliveira2026cautionary,schafer2024growing,o2026playful}.

Despite these challenges, research suggests that children can recognize and understand deceptive patterns \cite{sousa2023dark,renaud2024we,schafer2024growing,schafer2025if,arif2026understanding}. Renaud et al. \cite{renaud2024we}, for example, found that children aged 11–12 were aware of online deception, while Schäfer et al. \cite{schafer2024growing} found that many fifth-graders understood the intentions behind simple deceptive patterns. Research has also identified deceptive patterns in children's apps, including temporal, monetary, and psychological designs \cite{sousa2023dark}. However, recognizing deceptive patterns does not necessarily mean children can apply this understanding when interacting with manipulative interfaces. Research on persuasion knowledge suggests that although children develop an understanding of persuasive tactics, they may fail to apply it because of limited executive functioning and rapid processing of persuasive content \cite{friestad1994persuasion,rozendaal2011reconsidering}. Existing research on children and deceptive patterns has primarily examined which patterns children recognize, offering limited insight into what helps them apply this understanding when responding to manipulative interfaces.

\subsection{Children’s Online Privacy and the Role of Parents}

Children's understanding of online privacy has been studied extensively, 
particularly among elementary-school-aged children (roughly ages 5--11) 
\cite{kumar2017no,kumar2020strengthening,zhao2019make,zhang2016nosy}. 
Some of this work has examined how parents and children navigate privacy 
together \cite{kumar2017no,zhang2016nosy,wang2025they}. For example, 
Kumar et al.~\cite{kumar2017no} found that children aged 5--11 relied 
largely on their parents for privacy support, yet parents mostly mediated 
passively and deferred privacy conversations to the future. Similarly, 
Zhang-Kennedy et al.~\cite{zhang2016nosy} found that children aged 7--11 
and their parents held different threat models. While both worried about 
peers, media, and strangers, parents were also concerned about technology 
and children's own behavior, whereas children also viewed their parents 
as a potential threat.

Other studies have focused on how best to support children's privacy 
literacy \cite{kumar2020strengthening,chen2025empowering,zhang2017cyberheroes}. 
Kumar et al.~\cite{kumar2020strengthening} argue that privacy literacy 
should emphasize norms rather than rules, since children's judgments 
about disclosure depend on context, and that adults can use everyday 
learning moments to build this reasoning. Interventions have also targeted families directly. For example, an interactive ebook improved 7--9-year-olds' privacy knowledge and prompted longer parent--child privacy discussions than text-only materials \cite{zhang2017cyberheroes}, and short educational 
videos improved parents' security awareness, conversation strategies, 
and self-efficacy \cite{chen2025empowering}.

Finally, children appear more attuned to some privacy risks than others. 
Zhao et al.~\cite{zhao2019make} found that children aged 6--10 could 
identify and articulate risks such as oversharing information or revealing 
their real identities online, but were far less aware of risks such as 
online tracking and game promotions. Despite extensive research on children's understanding of privacy, on how  parents' and children's views of privacy differ, and on how to support children's privacy literacy, less is known about how children apply this understanding when an interface is designed to steer their choices, and  how their responses relate to the privacy guidance they receive at home. Prior work has linked restrictive parental mediation to children's limited 
understanding of implicit data practices \cite{desimpelaere2020children}, and has examined parents' own knowledge of deceptive patterns \cite{bessant2023exploring}, 
but to our knowledge, few studies have directly examined how different forms 
of parental mediation relate to how children recognize and respond to 
deceptive designs. As a result, it remains unclear whether recommended 
approaches to privacy education, such as emphasizing norms over rules, help children when they encounter manipulative interfaces \cite{kumar2020strengthening}.

\subsection{Types of Parental Mediation}

Warren \cite{warren2001words} defines ``parental mediation'' as ``any strategy parents use to control, supervise or interpret media content for children''. Research on parental mediation has identified several strategies parents 
use to guide children's media and internet use 
\cite{valkenburg1999developing,livingstone2008parental,nikken2014developing,bocking2009parental}.
\emph{Restrictive mediation} involves setting rules that limit children's 
media use, such as how much time they spend online, which apps or content 
they may access, and what activities they may engage in. \emph{Active 
mediation} involves talking with children about media, for example 
explaining content, discussing risks, or sharing and discussing their 
online activities. \emph{Co-use} refers to parents using media together 
with their children, without necessarily discussing it. \emph{Supervision} 
involves staying nearby or keeping an eye on children while they are 
online, whereas \emph{monitoring} involves checking children's online 
activity after use, such as reviewing browsing histories, messages, or 
contacts. Finally, \emph{technical mediation} relies on software or device 
settings, such as parental controls, filters, or app permissions, to 
restrict or track children's use.

Between roughly ages 9 and 12, children become increasingly autonomous in 
their media use and are less frequently supervised by parents 
\cite{bjelland2015associations}, yet this preteen period has received 
comparatively little attention in media research \cite{jordan2017growing}. 
As children gain more independence online, they are increasingly likely to encounter deceptive patterns without parental guidance and must draw on their own understanding to recognize them. However, research on parental mediation has largely focused on screen time, content exposure, and online risks such as stranger contact \cite{livingstone2008parental,nikken2014developing}. Understanding the types of parental mediation is crucial, as these strategies are most likely to continue supporting children when parents are no longer present to intervene.

\begin{figure}[htbp]
  \centering
  \includegraphics[scale=0.14]{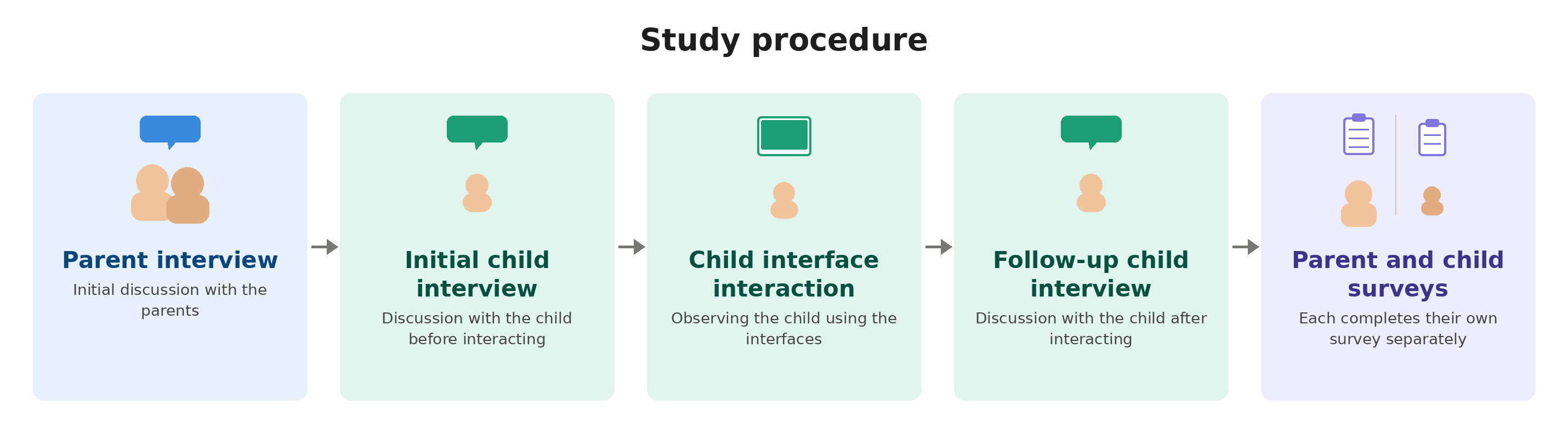}
  \caption{Study flow: parent interview, initial child interview, child interacts with the interfaces, follow-up child interview, then parent and child each complete their own survey.}
  \label{fig:research_flow}
\end{figure}

\section{Methodology}

Below, we elaborate on different components of the IRB-approved study, including participant recruitment, data collection, and analysis. Refer to Fig. \ref{fig:research_flow} to see the overall study flow.

\subsection{Participant Recruitment}
This study recruited children aged 8 to 12 and one of their guardians (aged 18 or older) to examine how guardians' privacy views relate to children's perceptions of deceptive patterns. Each family contributed one guardian and one child, both of whom had to be able to converse in English. All sessions were conducted in person in private rooms at a university or a public library in the U.S. Midwest between February and June 2026. Each family was compensated \$20 USD.

We recruited participants by distributing fliers through schools, libraries, social media, and direct email outreach, resulting in nine guardian--child pairs. Table~\ref{tab:demographics} summarizes participant characteristics. After the sessions, we emailed each guardian a short follow-up questionnaire about their child's school, asking for the school type (public or private), its approximate size, and whether the school had discussed online privacy with the child (\textit{Privacy Talk} in Table~\ref{tab:demographics}). All nine guardians responded. Our sample was drawn from the U.S. Midwest, and most guardians held professional or academic occupations; we discuss the implications of this in Section~\ref{sec:limt}.

\begin{table}[!h]
\centering
\small
\setlength{\tabcolsep}{3pt}
\renewcommand{\arraystretch}{1.1}
\resizebox{\textwidth}{!}{%
\begin{tabular}{l l l l l p{4.2cm} l l l l l}
\toprule
 & \multicolumn{3}{c}{\textbf{Child}} & \multicolumn{4}{c}{\textbf{Parent}} & \multicolumn{3}{c}{\textbf{Child's School}} \\
\cmidrule(lr){2-4} \cmidrule(lr){5-8} \cmidrule(lr){9-11}
\textbf{Pair} & \textbf{ID} & \textbf{Age} & \textbf{Gender} & \textbf{ID} & \textbf{Profession} & \textbf{Age} & \textbf{Gender} & \textbf{Type} & \textbf{Size} & \textbf{Privacy Talk} \\
\midrule
F1 & P1 & 12 & F & P10 & Small business owner and athletic coach & 44 & F & Private & Small  & Yes \\
F2 & P2 & 8  & F & P11 & University admin    & 40 & F & Private & Small  & No  \\
F3 & P3 & 11 & F & P12 & Psychiatrist        & 50 & F & Private & Small  & Yes \\
F4 & P4 & 10 & M & P13 & Professor           & 40 & F & Public  & Medium & No  \\
F5 & P5 & 10 & F & P14 & Professor           & 37 & F & Private & Small  & No  \\
F6 & P6 & 11 & M & P15 & Professor           & 43 & M & Public  & Small  & Yes \\
F7 & P7 & 11 & M & P16 & Lab manager         & 45 & F & Private & Medium & Yes \\
F8 & P8 & 9  & M & P17 & Lab staff           & 42 & M & Public  & Small  & Yes \\
F9 & P9 & 9  & F & P18 & PhD student         & 45 & F & Public  & Medium & Yes \\
\bottomrule
\end{tabular}%
}
\caption{Demographics of the nine parent--child pairs ($N = 18$). All families lived in the U.S. Midwest. \textit{Privacy Talk} indicates whether the parent reported that the child's school had discussed online privacy.}
\label{tab:demographics}
\end{table}

\subsection{Data Collection}
This study included three data collection components: semi-structured interviews with guardians and children, a walkthrough task in which children interacted with a prototype app, and a post-session survey. The interviews captured participants' experiences and perspectives, the walkthrough captured children's interactions and behaviors, and the survey captured participants' reflections. Together, these components enabled comparison of reported experiences and observed behaviors.
\begin{figure}[htbp]
  \centering
  \includegraphics[scale=0.10]{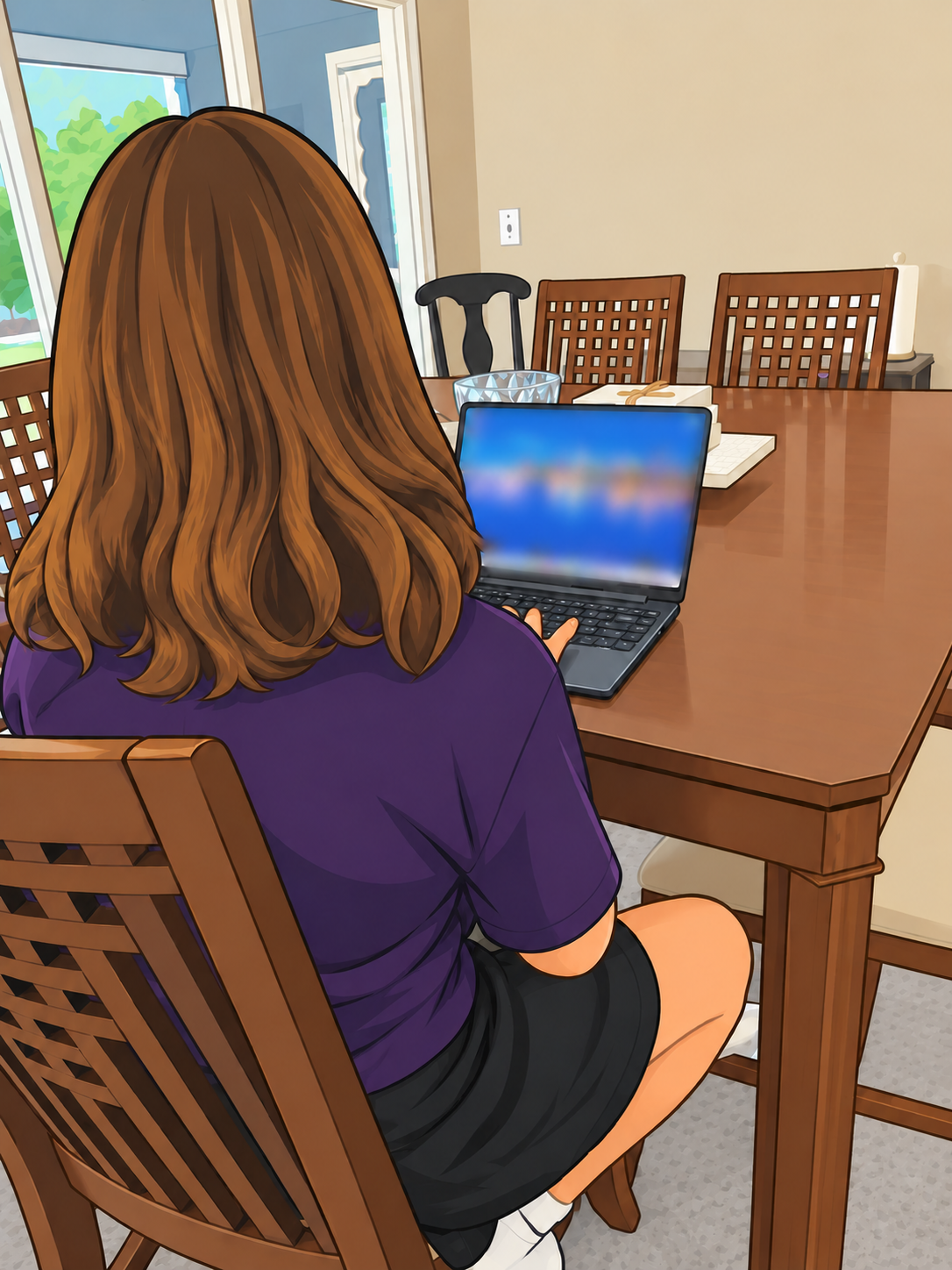}
  \caption{A child interacting with the walkthrough scenarios. The participant's appearance has been digitally transformed to protect their identity.}  
  \label{fig:pic}
\end{figure}

\subsubsection{Semi-structured Interviews}

Each session began with a semi-structured interview with the guardian, followed by a semi-structured interview with the child that took place before and after the walkthrough task. The guardian interview first collected background information and then explored the guardian's own views on privacy. We then asked whether their child used the internet or social media, why they allowed or restricted that use, and how they taught their child about online safety (see Appendix~\ref{appendix:parent}). The child interview was split around the walkthrough task. Beforehand, we asked children about their age and their use of technology and the internet, how they understood internet safety, and how they would handle potential online risks. Afterward, we asked about the parts of the interface they had just interacted with, and whether they could recall encountering similar deceptive patterns in real life, to gauge their understanding of manipulative design (see Appendix~\ref{appendix:child}).

All sessions were recorded in person using Zoom, which captured video, audio, and automatic captions. The recordings were then transcribed using Otter.ai, and the first and last authors reviewed each transcript against the original recording to correct errors before analysis.

\subsubsection{Scenario-Based Walkthrough}
To observe how children respond to deceptive patterns, the third and fourth authors built a clickable prototype of a fictional social media app in Figma, an interface design tool. The app was modeled on short-video platforms such as Instagram Reels and TikTok but used its own name and branding. Every element was clickable, and children used a mouse to interact with the prototype in Figma's prototype preview mode on a laptop, clicking and scrolling through a phone-sized frame as they would in a real app. Before starting, children were told that the app was not real. The prototype contained six scenarios, presented in a fixed order, each incorporating one or more deceptive patterns identified by Gray et al.~\cite{gray2018dark}, Mathur et al.~\cite{mathur2019dark}, Luguri and Strahilevitz~\cite{luguri2021shining}, and Brignull~\cite{Brignull_2023}. Together they spanned obstruction, nagging, sneaking (preselection), interface interference, and forced action (see Table~\ref{tab:dark-pattern-scenarios}).

\begin{table*}[htbp]
\centering
\small
\renewcommand{\arraystretch}{1.25}
\setlength{\tabcolsep}{.5pt}
\begin{tabular}{@{} >{\raggedright\arraybackslash}p{2.8cm}
                  >{\raggedright\arraybackslash}p{4.2cm}
                  >{\raggedright\arraybackslash}p{3.6cm}
                  >{\raggedright\arraybackslash}p{4.4cm} @{}}
\toprule
\textbf{Scenario} & \textbf{Description} & \textbf{Deceptive Pattern(s)} & \textbf{Successful Outcome} \\
\midrule
Location-sharing prompt & Required to share location to continue scrolling. & Obstruction (decline hidden); nagging (repeated prompt) & Found the hidden decline option and did not share location, including after repeated prompts. \\
\addlinespace
Notification request & Told that friends were waiting, encouraging the child to turn on notifications; the request repeats and the notification setting is preselected. & False social proof; nagging (repeated request); preselection (notifications on) & Declined notifications and recognized the nagging and the preselected notification setting. \\
\addlinespace
Public profile by default & Child's profile is publicly viewable by default. & Preselection (default privacy setting) & Noticed the public default and changed the profile to private. \\
\addlinespace
Digital wallet prompt & Prompted to add a digital wallet; the decline option is hard to find. & Obstruction & Found the decline option and did not add a wallet. \\
\addlinespace
Embedded shopping in Reels & Shopping and purchasing embedded within the Reels feed. & Preselection; obstruction; false urgency/scarcity; nagging & Did not complete a purchase, deselected any preselected items, and pointed out the false urgency. \\
\addlinespace
Avatar/profile creation & Encouraged to create avatars, complete profiles, and share publicly. & Confirmshaming; forced action (preselected public) & Created the avatar without sharing it publicly and pointed out the confirmshaming language. \\
\bottomrule
\end{tabular}
\caption{Simulated deceptive pattern scenarios and the criterion for a successful outcome in each. \textit{Partially successful}: the child recognized some but not all of the scenario's deceptive patterns. \textit{Unsuccessful}: the child recognized none of them.}
\label{tab:dark-pattern-scenarios}
\end{table*}

The location-sharing and digital wallet prompts test obstruction, and the location-sharing prompt also incorporates nagging. These scenarios examine whether children can exercise privacy and financial agency when refusal options are obscured or repeatedly presented. The notification request centers on false social proof, reinforced by nagging and a preselected setting, to examine emotional rather than purely structural manipulation, which has received less attention in research with younger users. The public profile default tests preselection, exposing children to privacy risks through passive defaults that require no action. The embedded shopping scenario combines multiple patterns, reflecting how commercial platforms often layer them, and adds urgency and financial stakes. Finally, the avatar/profile scenario combines confirmshaming with forced action in an identity-relevant context, where self-presentation may heighten the effect of manipulation. Together, the scenarios cover five pattern categories across varied contexts, allowing us to explore whether children's responses differed by pattern type or context. All interfaces were static and predefined, so every child encountered identical content.

The study recorded children's interactions with the interfaces for subsequent video analysis (see Section~\ref{sec:video_analysis}). This study used a think-aloud method, asking children to verbalize their observations, thoughts, and feelings while engaging with each scenario. This method provides insight into the cognitive processes underlying participants' task performance \cite{yang2003reconceptualizing}. Prior to deployment, members of the research team piloted the Figma interfaces to assess whether the deceptive patterns were identifiable as deliberate manipulations or design tricks. This manipulation check ensured the interfaces were realistic and subtle enough to approximate real-world app experiences without appearing as obvious tests that could alter children's behavior. Feedback was used to refine visual and interaction details (e.g., button placement, wording, timing) before the scenarios were finalized.

\subsubsection{Post-survey Questionnaire}
After the interviews and walkthrough task, guardians and children completed separate surveys. The research team developed both surveys specifically for this exploratory study. Rather than measuring established psychological constructs, the surveys were designed to complement the qualitative interviews and walkthrough by capturing participants' perceptions and self-reported experiences. The child survey focused on children's perceptions of the Figma interface immediately after interacting with it, while the guardian survey explored guardians' confidence in addressing their child's online privacy and their self-reported engagement in privacy-related practices. Items used a 5-point Likert scale. The survey items were not adapted from or validated against existing instruments. Given their exploratory purpose, the lack of instrument validation, and the small sample size, we report survey responses descriptively rather than treating them as validated measures of established constructs. See Appendix~\ref{appendix:survey} for the survey questions.

\subsection{Data Analysis}
This study used two types of qualitative analysis. The first was a thematic analysis of the semi-structured interviews with guardians and children, and the second was a directed content analysis of video recordings of children's interactions with the interface. This dual approach allowed the study to triangulate self-reported attitudes and beliefs from the interviews against observed behavioral patterns from the video-recorded interactions, surfacing any gaps between what guardians perceive or hope for and what children actually do when using the interface.

\subsubsection{Semi-structured interview}

We analyzed the parent and child interview transcripts using reflexive thematic analysis~\cite{braun2006using, clarke2017thematic}. The first author familiarized themselves with the data by reading each transcript in full, then coded passages relevant to our research questions inductively and grouped related codes and their supporting quotes into candidate themes and sub-themes. These were organized in a hierarchical mind map using XMind to examine how codes and themes related to one another. The second and last authors then reviewed each candidate theme alongside its supporting quotes. Over multiple discussions, all three authors decided which themes and quotes to retain, merge, or revise, and agreed on the definition and name of each final theme. We reached consensus through discussion rather than measuring inter-coder agreement, consistent with the reflexive approach, and we treated themes as constructed through our interpretation rather than as emerging directly from the data.

\subsubsection{User Interface Interaction} \label{sec:video_analysis}

\begin{table}[htbp]
\centering
\small
\renewcommand{\arraystretch}{1.25}
\begin{tabular}{@{} l p{10cm} @{}}
\toprule
\textbf{Style} & \textbf{Definition} \\
\midrule
Active & The child intentionally participates, creates, or makes decisions rather than simply consuming content. \\
\addlinespace
Passive & The child consumes content without actively interacting or contributing. \\
\addlinespace
Interactive & The child engages in two-way exchanges with the interface. \\
\addlinespace
Multitasking & The child performs multiple activities at once (e.g., talking while scrolling). \\
\bottomrule
\end{tabular}
\caption{Behavior styles coded in the walkthrough videos, adapted from Chi and Wylie~\cite{chi2014icap} and Ophir et al.~\cite{ophir2009cognitive}.}
\label{tab:behavioral-types}
\end{table}

To examine children's interactions with the prototype, we manually analyzed the walkthrough video recordings using directed content analysis~\cite{hsieh2005three}, which applies predefined codes drawn from existing theory to the data. These observations addressed RQ1 by examining how children responded to privacy- and safety-related interface elements and navigated potentially manipulative designs. The first author coded all nine recordings using the predefined codes and success criteria defined in Tables~\ref{tab:behavioral-types}, \ref{tab:taxonomy}, and \ref{tab:dark-pattern-scenarios}. Rather than having a second coder independently code a subset of recordings to calculate an inter-rater reliability statistic, the second author reviewed all coded segments across the nine recordings against the original video recordings, assessing the appropriateness of the assigned codes. Given the small dataset, this full review was feasible and allowed the authors to discuss and resolve disagreements at the segment level. This approach was particularly useful for interpretive codes, such as hesitation or expressions of concern, whose meaning often depended on the child's concurrent think-aloud explanations and the surrounding interaction.

To characterize participants' interactions more systematically, we analyzed their behavior at two levels of abstraction, which were behavior style and behavior type. Behavior style captured the general mode of engagement exhibited by participants, whereas behavior type captured the specific activity or goal participants were pursuing during their interaction with the application. The behavior styles were adapted from Chi and Wylie \cite{chi2014icap} and Ophir et al. \cite{ophir2009cognitive}. These behaviors include active, passive, interactive, and multitasking behaviors (see Table \ref{tab:behavioral-types}).

\begin{table}[htbp]
\centering
\small
\renewcommand{\arraystretch}{1.25}
\begin{tabular}{@{} >{\raggedright\arraybackslash}p{2.2cm}
                  >{\raggedright\arraybackslash}p{10.5cm} @{}}
\toprule
\textbf{Behavior} & \textbf{Definition} \\
\midrule
Exploring & Open-ended browsing of the interface (e.g., scrolling the feed, tapping around a screen) without a specific goal. \\
\addlinespace
Learning & Reading on-screen text or options to understand what a prompt or feature does before acting. \\
\addlinespace
Comparing & Weighing the available options on a screen (e.g., accept versus decline, items in the shop) before choosing. \\
\addlinespace
Deciding & Committing to a choice on a prompt, such as allowing or declining location sharing, notifications, or adding a wallet. \\
\bottomrule
\end{tabular}
\caption{Behavior types coded in the walkthrough videos, adapted from \cite{broder2002taxonomy, pirolli1999information, o2008user}.}
\label{tab:taxonomy}
\end{table}

To characterize children's engagement, we began with ten behavior categories (exploring, searching, learning, comparing, deciding, creating, communicating, monitoring, returning, and exiting) drawn from Broder's web search intent categories~\cite{broder2002taxonomy}, information foraging theory~\cite{pirolli1999information}, and O'Brien and Toms' user engagement framework~\cite{o2008user}. Because the walkthrough was a single session with a fixed set of screens, several categories (e.g., searching, communicating, returning) could not occur, so we retained the four observed in the videos and adapted their definitions to the prototype (Table~\ref{tab:taxonomy}).

The study recorded six measures, which were task completion time, success rate, button clicks, user reactions, behavior style, and behavior type. Task completion time measured the duration of each scenario. Success in each scenario was coded as successful, partially successful, or unsuccessful. An outcome was \textit{successful} when the child recognized all of the deceptive patterns in the scenario and met its success criterion (Table~\ref{tab:dark-pattern-scenarios}), \textit{partially successful} when the child recognized some but not all of the patterns, and \textit{unsuccessful} when the child recognized none of them. Button clicks recorded interactions with clickable interface elements. User reactions captured participants' verbal and behavioral responses, including comments, hesitation, and expressions of concern. Behavior style classified participants' overall engagement with the interface as active, passive, interactive, or multitasking (Table~\ref{tab:behavioral-types}). Behavior type captured the specific actions participants took within each scenario, such as deciding, learning, exploring, or comparing (Table~\ref{tab:taxonomy}).

\section{Results}
\label{results}
In this section, we report findings from the parent and child interviews, the children's scenario-based walkthroughs, and the post-surveys. Sections~\ref{sec:family_context}--\ref{sec:limits} report findings from the parent interviews: parents' views on privacy, how they mediated their children's technology use, and the limits they described in that mediation. Section~\ref{sec:child_responses} reports findings from the children's scenario-based walkthroughs and interviews. Section~\ref{sec:paired} then brings the two together, placing each parent's account alongside their own child's, and Section~\ref{sec:survey} reports descriptive post-survey results. Children are P1--P9 and parents are P10--P18; each child P$n$ is paired with parent P$(n+9)$.

\subsection{Parents' Views on Privacy}
\label{sec:family_context}
We begin with findings from the parent interviews. Because we examine how parents' own understanding of privacy relates to their children's responses, we first describe how parents viewed privacy. Three views recurred: privacy as control over personal information, awareness of how interfaces collect and use data, and concern shaped by past incidents. We return to how these views relate to parents' mediation and their children's recognition in \S\ref{sec:paired}.

\subsubsection{Privacy as Control over Personal Information}
Parents described privacy as controlling what personal information they disclosed, both online and offline. P13 explained, \textit{``I try not to share anything with others, and if I receive mail, before I throw it away, I'll tear off the personal information, like my name and address… and online, I'll use my own important passwords rather than letting my computer remember them.''} Several parents extended this control to decisions made on their children's behalf. P14 said, \textit{``It is important to me that I am aware how my data is being collected and why, what it might be used for, and what third parties might have subsequent access to it down the line. It's particularly important to me that I make deliberate choices about her privacy, given that she is not necessarily able to make those choices for herself.''} P15 applied this to photos of their children: \textit{``I don't share pictures of the kids with their names attached. We make sure that there's no, on all of our photos, there's no location, there's no geotag, so there's the location attached to photos.''} P17 framed limiting what they posted as protecting their children's futures: \textit{``The less we put on there now, the less they have in the future to worry about. So we don't want to you know mess up their possible future by putting a bunch of stuff on there that you know they don't even know about''.}

\subsubsection{Awareness of Interface Data Practices}
Parents also described noticing interfaces and services that collected or used their information in ways that felt unclear or excessive, the experiences closest to the deceptive patterns in our walkthrough. P10 stated, \textit{``feels like, you know, you get all those if you play one game on your phone and you get a million apps for all the other games, or you get a million ads for all the other games, and like those I try to stay away from, because those seem pretty shady too.''} P12 described, \textit{``Very cautious, I really, but I don't have any particular mind as I do my stuff, I just know that there's a chance that you know things can be because it's weird, I noticed when, when I, if I check on something, or maybe like we were, we applied for a HELOC, a kind of loan, and then after we got approved, I started getting emails, you know, like adverts, adverts things about HELOC, and I'm just like, how did they know that?''} P16 connected online shopping to the resale of their data: \textit{``I heard that when shopping on some of the Asian grocery websites, like Weee or Yummy... they sell your information to third parties to get profit from it, so I do notice that after I shop on those websites, I got like scam calls.''}

Awareness did not always lead to caution. P18 described knowingly trading personal information for benefits: \textit{``Some application needs my privacy, and they will give me some benefit. For example, can I can use some free service, or can I can get some free product? I just give my personal information without any careful thinking.''} For P18, data collection was an acceptable exchange rather than a risk to manage.

\subsubsection{Concern Shaped by Past Incidents}
Many parents' views of privacy were shaped by incidents they or people they knew had experienced, including identity theft, fraudulent claims, data breaches, impersonation, and scams. P11 recounted that \textit{``Once, during COVID, somebody filed for unemployment using, I think, my Social Security number, so I had to go through the process of reporting it and dealing with it.''} P12 described an incident in which someone used their identity to contact a patient about medical billing. Parents described responding with protective measures such as freezing credit reports, subscribing to identity theft protection, and verifying suspicious communications. These incidents framed privacy risk as explicit threats from bad actors, such as scams that could be recognized and deleted. P18 described such content as \textit{``some sexual thing or some scam they need some money but I just erased them.''} Fewer parents described the design of everyday apps as a privacy risk in itself.

\subsection{Parental Mediation}
\label{sec:mediation}

To understand how parents' privacy views translated into practice, we examined how they mediated their children's technology use. We draw on parental mediation research~\cite{livingstone2008parental}, which distinguishes active mediation, in which parents talk with children about online content and risks, from restrictive mediation, in which parents set rules and limits, and from technical and monitoring approaches. Parents in our study typically combined several of these strategies. We first describe active mediation, including both parent-initiated teaching and child-initiated dialogue, then restrictive and monitoring mediation, and finally limited mediation, in which conversations about privacy remained minimal or were replaced by rules.

\subsubsection{Active Mediation}
Many parents described proactively teaching their children basic safety and privacy behaviors to help them recognize and respond to potentially harmful situations online. Their discussions focused on helping children distinguish advertisements from legitimate content, avoid clicking suspicious links, recognize trustworthy websites, and understand the financial consequences of online purchases and subscriptions. For example, P15 stated, \textit{``We make it clear that you should never click links on websites. We especially try to distinguish between advertisements and the actual page content. We also teach them to pay attention to the domain in the URL. If something pops up unexpectedly, don't ever click on it''}. These practices demonstrate how parents sought to build children's digital literacy alongside basic online safety behaviors.

Other parents, such as P12 and P13, used real-world incidents as teachable moments to discuss online safety. After P12's daughter repeatedly followed a peer who posted furry-related content, P12 asked why she kept watching content she claimed not to like: \textit{``Why are you [...] spending enough time following this person because you're always watching what the person posts?''} Although her daughter initially did not see a problem with this behavior, P12 used the situation to show how repeatedly monitoring others' online activity can consume one's mental space. P13, in contrast, learned of a risk only after their son had encountered it. While using YouTube at school, he clicked a pop-up link inviting him to join a club. As P13 recounted, \textit{``he said he just sense it is not right, so he just closed that window [...] After that, I just told him not every link you can click, because you know some, some will just get all your information off [...] so you'd rather be wise to use those to click those link.''} P13 then used this moment to explain how clicking unfamiliar links can lead to stolen information, encouraging their son not to click on things he does not recognize.

Other participants drew on their professional knowledge and workplace experiences to educate their children about online privacy and safety. For example, P14 described incorporating discussions from their undergraduate lectures into conversations with their child, explaining the risks associated with different applications and websites while encouraging questions about technology use. P17 explained that cybersecurity incidents encountered through work became opportunities to discuss online threats and reinforce safe online behaviors with their children, saying, \textit{``Yeah... she (wife) works in IT... somebody got a virus... and shut the whole college into chaos... so she'll talk to the kids about that kind of stuff... this is why we have to... not open something you don't know what it is... she definitely is part of and drives some of it too.''}

Beyond proactively teaching their children about online privacy, some parents encouraged their children to ask questions and consult them when they were uncertain. As P14 stated, \textit{``She's welcome to search for things and watch things and learn about the world, and we talk about it.''} P15 similarly kept an open door for questions, telling their children, \textit{``if you're not sure what to do, just step back and ask for us, ask for our help.''} P15 also talked through information requests with their children as they arose: \textit{``if we're signing them up for something, and we're filling out a form or giving information, we explain why it's reasonable to give this information in this context, whereas it's like we're signing up for game board game arena or something like that, some gaming site [...] We explain that, like, there's no reason for this like entity to need these details, so we're not going to give them like our information.''}

Parents also noted that this dialogue depended on whether children themselves were willing to start it. P16 contrasted their two sons: \textit{``I worry about [my first son] more, because [my first son] does not communicate me very well, but [my second son], he will give a talk, he will say to ask things, so I think it's easier if it kids can, you know, communicate with you all the time.''} Child-initiated dialogue thus depended not only on parents' openness but also on children's willingness to raise what they encountered, leaving parents with less insight into children who rarely shared.

\subsubsection{Restrictive and Monitoring Mediation}

Even though many parents described teaching proactively and keeping communication with their children open, many also practiced restrictive mediation, particularly around social media. P15 stated, \textit{``I think they have no autonomy, and should have no autonomy with lifestyle media, but that's not a thing, that's been like, I don't want that to exist in their world until they're older, until they have a like clear sense of self, and so that's that's there's no balance there.''} P10 similarly explained, \textit{``I don't think she is responsible for it, and also I don't want to invite that into our home for not knowing who she's talking to or how they're, how they're talking.''} P13 shared this sentiment: \textit{``He doesn't have social media, because I just want to keep him away from social media.''} At the most restrictive end, P18 strictly limited their children's smartphone use: \textit{``Because I am strictly prohibited using their smartphone.''} P18 also hoped to delay social media use until college.

While most parents restricted social media, all of them monitored their children's internet use in some form. P12 combined monitoring with guidance. When her children shopped online, she and her husband asked them to check that sites were legitimate and to share the sites with them before buying: \textit{``you have to be careful, check the reviews on the site to make sure that it's not a fictitious company [...] they should make sure they share with him where they, what site it is before [...] [they] buy things on non-reputable sites.''} Other parents relied on technical tools and household arrangements. P15 described layered monitoring: \textit{``I get a weekly email with search terms, the your the websites they went to, and the kind of risk rating of the different sites. We also, I mean, block lots of domains and sites [...] based on risk, and like the age rating, essentially of the website content.''} P15 paired these tools with rules about device use: \textit{``their computers are in adjacent to our family room, so we can all see the screens, and then they don't have unrestricted use of their iPads [...] They get asked permission every time they get on their iPads, and we need to know what they're up to, and check the screen time regularly.''} P14, by contrast, used lighter-touch controls: \textit{``I don't monitor her every interaction, but I do use permissions to fulfill what apps are available to her and we, I suppose, also have a rule that her iPad is not in her room, so it's in used in common places in the house.''} Yet parents recognized that neither rules nor tools could cover everything. As P13 noted, \textit{``I cannot sit with him all the time, and he's very sneaky, so sometimes he just do things without my permission.''} Restrictive and monitoring mediation thus helped parents limit children's exposure, but left children to handle much of what they encountered on their own.

\subsubsection{Limited Mediation}
Not all parents discussed privacy with their children in depth. For some, conversations stayed at a baseline level, focused on explicit risks such as intimate images rather than everyday privacy decisions. P11 said, \textit{``We haven't talked a lot about privacy online, but we have had baseline discussions about, like, again, like, not taking pictures with your clothes off [...] trying to sort of make them aware of gross behavior without scaring them. Well, we really haven't, like, we haven't gotten as far as talking about online things, because they don't spend a lot of time online.''} Other parents acknowledged that they had discussed privacy less than they felt they should. When asked whether they discussed online privacy and safety with their child, P16 reflected, \textit{``That's a very good question. I should, but I just. the only thing I did is just do not like pay for anything, doing if you want to buy something, definitely let me know ahead of time.''}

For some parents, restriction appeared to stand in for conversation. Because P18's children had no social media applications, P18 saw little reason to discuss them: \textit{``They don't have any related application on their phone, so I never think about that kind of things.''} P18 likewise did not discuss routine prompts such as in-app advertisements: \textit{``Not really, because sometimes they are using Duolingo, and there are a lot of gaming advertising pop up, but I don't care because I know that kind of pop up things. Not asking for sharing their personal information.''} In these families, guidance addressed explicit dangers or was replaced by rules, leaving everyday interface prompts, such as advertisements, pop-ups, and requests for information, largely undiscussed.

\subsection{Limits of Mediation}
\label{sec:limits}
Although parents used a range of mediation strategies, they also described barriers that limited how far these strategies could protect their children.

\subsubsection{Feeling Powerless against Technology}
Parents described the complexity and opacity of digital technologies as limiting their ability to understand and protect their children's privacy. P15 explained, \textit{``I worry about the ways in which she's digitally profiled. She's creating records and digital footprint that are going to be used to make decisions about her down the line… I am very concerned about how algorithmic decision making is going to influence her future…''} P17 expressed uncertainty about emerging risks: \textit{``...with the rise of AI and the images they can generate and stuff like that... that's going to be a concern for like blackmailing... Everything's searchable online... at some point that could come into play.''} Parents also expressed uncertainty about platforms marketed to children. P11 noted, \textit{``hearing horror stories, like Roblox... the way that parents have thought that it's safe for kids and there are just predators lurking. I mean, any sort of environment where it's not clear who they're interacting with, it makes me really nervous.''} These risks lay largely outside what parents' rules and conversations could address.

\subsubsection{Limited Insight into Children's Online Behavior}
Parents also described limited insight into what their children did and understood online. P16 said, \textit{``The thing I learned is that you cannot trust them 100\% because they lie to you. They lie for all kinds of reasons. One time, I trusted my first son to attend his online Chinese class, but he was watching YouTube videos instead. ... That's why I started turning off the Wi-Fi when he was supposed to be doing his schoolwork.''} As noted above, P16 also found it harder to guide a son who rarely communicated. Other parents were unsure how much their children understood. P13 said, \textit{``...I just feel like he doesn't have any awareness about privacy and online phishing or other things... you cannot share your privacy information, and you cannot click the link... so you should be just a bit careful about that.''} P12 similarly explained, \textit{``I'm not so confident, because at this point I just worry about, worry about how much I what what she knows is safe. What she accepts, you know.''}

\subsubsection{Views on Children's Autonomy}
Parents differed in how much autonomy they believed children should have online, which shaped how far they relied on restriction. As described above, P10, P13, P15, and P18 preferred to keep social media out of their children's lives for now. P13 explained this in terms of maturity: \textit{``I just feel he is so young to use social media because he is not like mature enough to tell the right and wrong things, and I just feel like those influencers, he can just easily just be influenced by them, but he doesn't know if it's right or not...''} Others saw autonomy as something to build gradually. P17 explained, \textit{``By the time I get like 14 to 16, I think probably start open those things. They get a lot more peer pressure by then too... we'll probably have to start, you know, saying okay, well, you know, watch out for these kinds of things... or put these kind of limitations in place, and kind of start ramping up by then.''} P17 compared this process to learning to drive. P14 framed the balance as protection without censorship: \textit{``I think what the internet is positive for is a source of information… I'm not interested in censorship, I'm interested in keeping her safe, but not silent. [...] the balance is in preventing the profiling and data collection and engagement with social media platforms that are not age appropriate.''} Parents' views thus ranged from delaying social media until adolescence or later to introducing independence gradually with guidance.

\subsection{How Children Responded to Deceptive Patterns}
\label{sec:child_responses}

We now turn from the parent interviews to children's interactions, drawing on their think-aloud comments and interviews. Children varied in how far they recognized the deceptive patterns across the six scenarios (Table~\ref{tab:task_summary}). P1, P3, P5, P7, and P8 recognized the deceptive patterns in all six scenarios. P4 recognized them in five scenarios and partially in the notification scenario, and P6 recognized them in five scenarios but not in the final scenario. P9 recognized the deceptive patterns in three scenarios and partially in the remaining three. P2 progressed through every scenario but did not recognize the deceptive patterns in any of them. For the paired analysis (Table~\ref{tab:dyadic analysis}), we grouped these outcomes into three recognition levels: \textit{High} (at least five scenarios fully recognized; P1, P3--P8), \textit{Moderate} (P9), and \textit{Low} (P2).

\begin{table}[htbp]
\centering
\resizebox{\textwidth}{!}{%
\begin{tabular}{lccccll}
\toprule
\textbf{Participant} & \textbf{Avg. Time (min:sec)} & \textbf{Scenario Recognition Outcomes} & \textbf{Avg. Buttons Clicked} & \textbf{Dominant Behavior Style} &  \textbf{Top Behavior Types (count)} \\
\midrule
P1 & 1:19 & 6Y & 5.83 & 6 Active / 6 Interactive & Deciding (6), Learning (5), Exploring (3) \\
P2 & 1:21 & 6N & 5.33 & 6 Passive / 6 Interactive & Deciding (3), Exploring (2) \\
P3 & 1:13 & 6Y & 6.00 & 6 Active / 6 Interactive & Deciding (6), Comparing (5), Learning (3) \\
P4 & 2:35 & 5Y / 1P & 5.17 & 6 Active / 6 Interactive & Deciding (6), Exploring (4), Learning (2) \\
P5 & 2:11 & 6Y & 5.50 & 6 Active / 6 Interactive & Deciding (5), Comparing (4), Learning (3) \\
P6 & 2:27 & 5Y / 1N & 3.67 & 6 Active / 6 Interactive & Deciding (6), Exploring (4), Learning (3) \\
P7 & 2:22 & 6Y & 6.17 & 6 Active / 6 Interactive & Deciding (6), Comparing (5), Learning (3) \\
P8 & 2:38 & 6Y & 6.00 & 6 Active / 6 Interactive & Deciding (6), Comparing (6), Learning (3) \\
P9 & 1:40 & 3Y / 3P & 3.67 & 6 Active / 6 Interactive & Deciding (6), Exploring (3), Learning (2) \\
\bottomrule
\end{tabular}%
}
\caption{Participant-Level Summary Across All Six Tasks}
\label{tab:task_summary}
\end{table}

Recognition did not track speed or amount of interaction. P3 had the shortest average time (1:13) and P8 the longest (2:38), yet both recognized the deceptive patterns in all six scenarios, and P2 averaged 5.33 button interactions despite recognizing none. What differed was the kind of engagement. P2 was the only child whose behavior was coded as passive throughout and whose coded behaviors were mostly \textbf{Deciding} and \textbf{Exploring}. Several children who recognized every deceptive pattern also frequently showed \textbf{Comparing} and \textbf{Learning} alongside \textbf{Deciding}. Below, we describe four ways children responded to the deceptive patterns, drawing on their think-aloud comments and interviews.

\begin{figure*}[t]
  \centering
    \begin{minipage}[t]{0.31\textwidth}
    \centering
    \includegraphics[width=\textwidth]{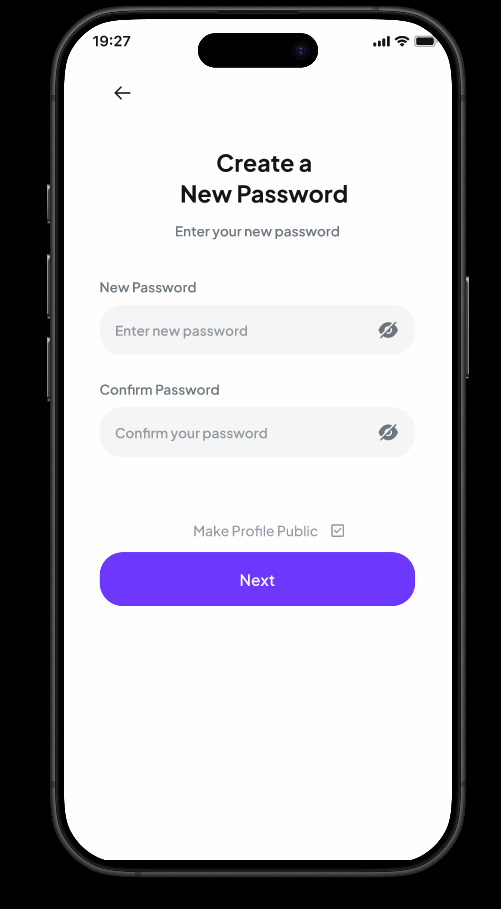}
    \caption{The small, low-salience “Make Profile Public” option can make a consequential privacy decision easy to overlook. For children, defaulting or encouraging a public profile may expose personal information and interactions to a broader audience than they anticipate.}
    \label{public}
  \end{minipage}
    \hfill
  \begin{minipage}[t]{0.35\textwidth}
    \centering
    \includegraphics[width=\textwidth]{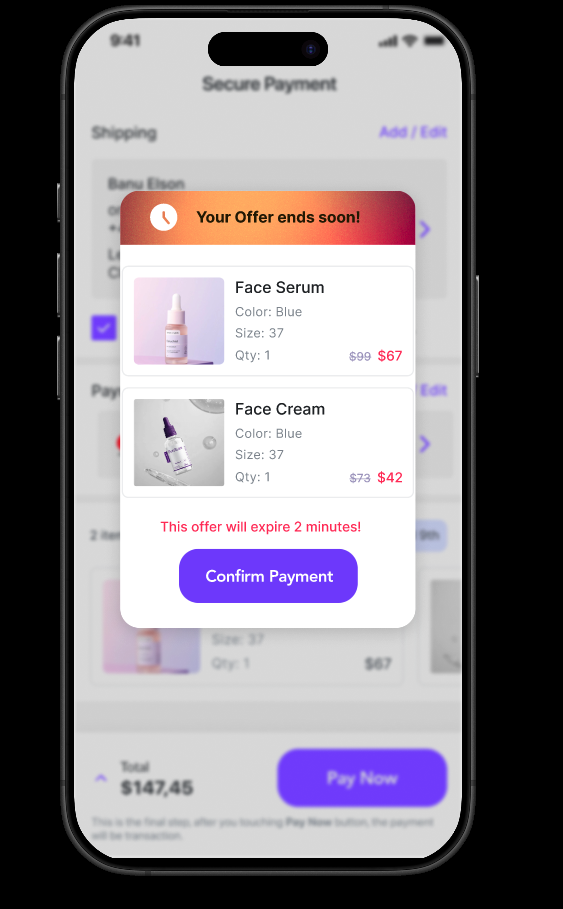}
    \caption{The countdown timer and ``Your Offer ends soon!'' message create a sense of fake urgency, pressuring users to complete a purchase before they have time to reconsider. By suggesting that the discounted prices will disappear within two minutes, the interface encourages impulsive decision-making rather than informed choice.}
    \label{fake_urgency}
  \end{minipage}
  \hfill
  \begin{minipage}[t]{0.30\textwidth}
    \centering
    \includegraphics[width=\textwidth]{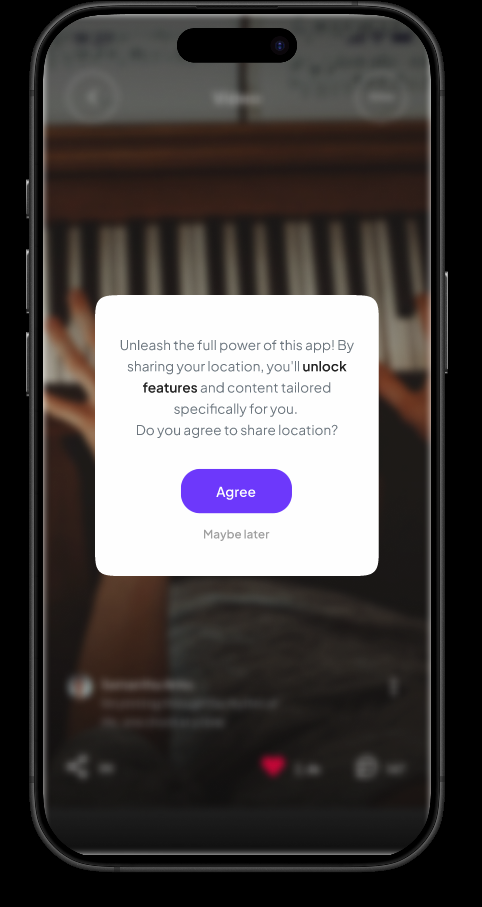}
    \caption{Obstruction: The more prominent and noticeable “Agree” button draws users’ attention toward accepting location sharing, while the less noticeable ``Maybe later'' option makes declining less apparent.}
    \label{obstruction}
  \end{minipage}
\end{figure*}

\subsubsection{Noticing Visual Steering}
Many children navigated the obstruction pattern by reading the prompt carefully (see Fig.~\ref{obstruction}), and some went further to name how the design steered them. In the third scenario, where children were asked to make their profile private, P7 noticed both that the application was encouraging a public profile and that the option to keep it private was small: \textit{``Oh, wait, the `Yes' is, like, really, really small. Well, no, it's blending in with the white background, and then the `Later' has a really big blue [button].''} P8 recognized similar visual emphasis in the fourth scenario, where they were asked to opt out of the wallet: \textit{``Um, this one is the one that they're trying to make you do because it's highlighted, kind of, and this one's darker.''}

\subsubsection{Interpreting Consequences}
Beyond noticing salient features, some children reasoned about what a choice would lead to. In the third scenario, P7 recognized that a public profile would be visible to everyone before the profile had even been created (see Fig.~\ref{public}): \textit{``yeah, this make profile public thing, because now maybe everyone can access your profile.''} In Scenario 6, P5 questioned adding a profile picture and noticed that the interface offered no way to decline: \textit{``It wants you to add a picture of yourself, which is like not good, because then it knows like what you look like, and they could track you down, and it doesn't let you say no, it just says okay, it only lets you say okay.''} In the shopping scenario (see Fig.~\ref{fake_urgency}), P8 drew on their own past experience to question a time-limited offer: \textit{``It says your offer ends soon, but that's not that much, really. And like sometimes offers are actually fake because I've done that on Roblox before, like it said this offer will expire in two hours. I came back the next day; it was still there.''}

Some children also explained the data practices behind these designs. P7 described a data broker as follows: \textit{``they're people who want just like one to take your data, and I don't know, they take your data, and then they like they can sell to like, more to, so like scam callers can target you''}. P4 described knowing how to erase their search history to protect their privacy.

\subsubsection{Recognizing the Request but Not the Design}
Some children understood what an application was asking for and declined, without recognizing how the interface shaped the choice. In Scenario 1, P9 noticed that the message was asking for their location and chose ``Maybe later,'' explaining, \textit{``Because a stranger can come, because strangers can come to our house.''} However, P9 did not recognize that options on the following screen had been preselected. In Scenario 2, P9 again recognized that the application wanted them to enable notifications and declined, explaining, \textit{``Because it's noisy.''} P9 did not, however, recognize the preselected notification settings. In Scenario 3, P9 did not notice that ``Make profile public'' was preselected and displayed less prominently (see Fig.~\ref{public}), although they did find how to set the profile to private on the main page. Across these scenarios, P9 understood the purpose of each request but did not identify the design features that steered their choices.

\subsubsection{Following the Most Prominent Option}
P2 tended to select the most prominent option before reading the prompt. In Scenario 1, P2 immediately clicked ``Share location'' and, when asked what sharing their location meant, could not explain it. P2 likewise immediately enabled notifications in Scenario 2, clicked ``Confirm payment'' in Scenario 5 (where the prompt was designed to reappear repeatedly), and clicked ``Share profile picture'' in Scenario 6. P2 did not recognize any preselected options or find how to make their profile private in Scenario 3. P2 also had difficulty with concepts such as credit cards and digital wallets, suggesting that their difficulties involved unfamiliar concepts within the scenarios as well as the interface manipulations themselves.

\subsection{Paired Parent and Child Accounts}
\label{sec:paired}

Bringing together the parent interviews and children's walkthroughs, Table~\ref{tab:dyadic analysis} places each parent's mediation, using the categories in \S\ref{sec:mediation}, alongside their child's recognition of deceptive patterns. We characterized talk depth from parents' interview accounts: \textit{Low} for rules or baseline safety only; \textit{Limited} for occasional talk prompted by incidents or the child's questions; \textit{Moderate} for recurring talk about online risks; and \textit{High} for regular, deliberate explanation of data practices. Below, we also draw on parents' views of privacy (\S\ref{sec:family_context}) to consider how those views, parents' mediation, and children's recognition related within each family. With nine dyads, we describe patterns we observed across pairs rather than effects of particular mediation styles.

\begin{table}[!h]
\centering
\footnotesize
\setlength{\tabcolsep}{3pt}
\renewcommand{\arraystretch}{1.2}
\begin{tabular}{l c p{2.6cm} c c p{3.6cm}}
\toprule
\textbf{Child (Parent)} & \textbf{Talk Depth} & \textbf{Mediation} & \textbf{School Talk} & \textbf{Recognition} & \textbf{Example} \\
\midrule
P1 (P10) & Limited  & Active; Restrictive and Monitoring & Yes & High     & Beware of what they post online and not everything online are true.\\
\midrule
P2 (P11) & Low      & Limited & No  & Low & Baseline discussions about not taking pictures with clothes off.\\
\midrule
P3 (P12) & Moderate & Active; Restrictive and Monitoring & Yes & High    & Has a conversation whenever daughter brings up something alarming.\\
\midrule
P4 (P13) & Limited  & Active (incident-driven); Restrictive and Monitoring & No  & High     & When child encounters a privacy issue, instructs child on the danger of it. \\
\midrule
P5 (P14) & High     & Active; Restrictive and Monitoring & No  & High     & Takes child to college privacy lectures. \\
\midrule
P6 (P15) & High     & Active; Restrictive and Monitoring & Yes & High     & Does it every time the child downloads an application and explains the data cycle.\\
\midrule
P7 (P16) & Limited  & Limited & Yes & High     & Asks questions when purchasing items or asks questions if they do not know something. \\
\midrule
P8 (P17) & High     & Active & Yes & High     & Taking child to Cub Scouts where they educate them on privacy.\\
\midrule
P9 (P18) & Low      & Restrictive and Monitoring & Yes & Moderate & Parents ask them to avoid unfamiliar things. \\
\bottomrule
\end{tabular}
\caption{Parental Mediation Characteristics and Children's Deceptive Pattern Recognition by Parent--Child Pair. Mediation categories follow \S\ref{sec:mediation}; Active includes both parent-initiated teaching and child-initiated dialogue.}
\label{tab:dyadic analysis}
\end{table}

\textbf{Active mediation carried parents’ views into conversations about apps.} In several pairs, the privacy view a parent described appeared in how they talked with their child and in how the child reasoned during the walkthrough. P15 limited the information they shared with applications, described explaining the data cycle to P6 each time the child downloaded an application, and monitored their children's browsing through weekly reports; P6 recognized the deceptive patterns in five of six scenarios. P14 emphasized knowing how data is collected and who might access it, brought P5 to their undergraduate privacy lectures, and restricted app permissions; P5 reasoned that a profile picture would let others know \textit{``what you look like, and they could track you down.''} In both pairs, the parent's active mediation addressed how apps collect information, not only which actions to avoid.

\begin{figure}
    \centering
    \includegraphics[width=1.0\linewidth]{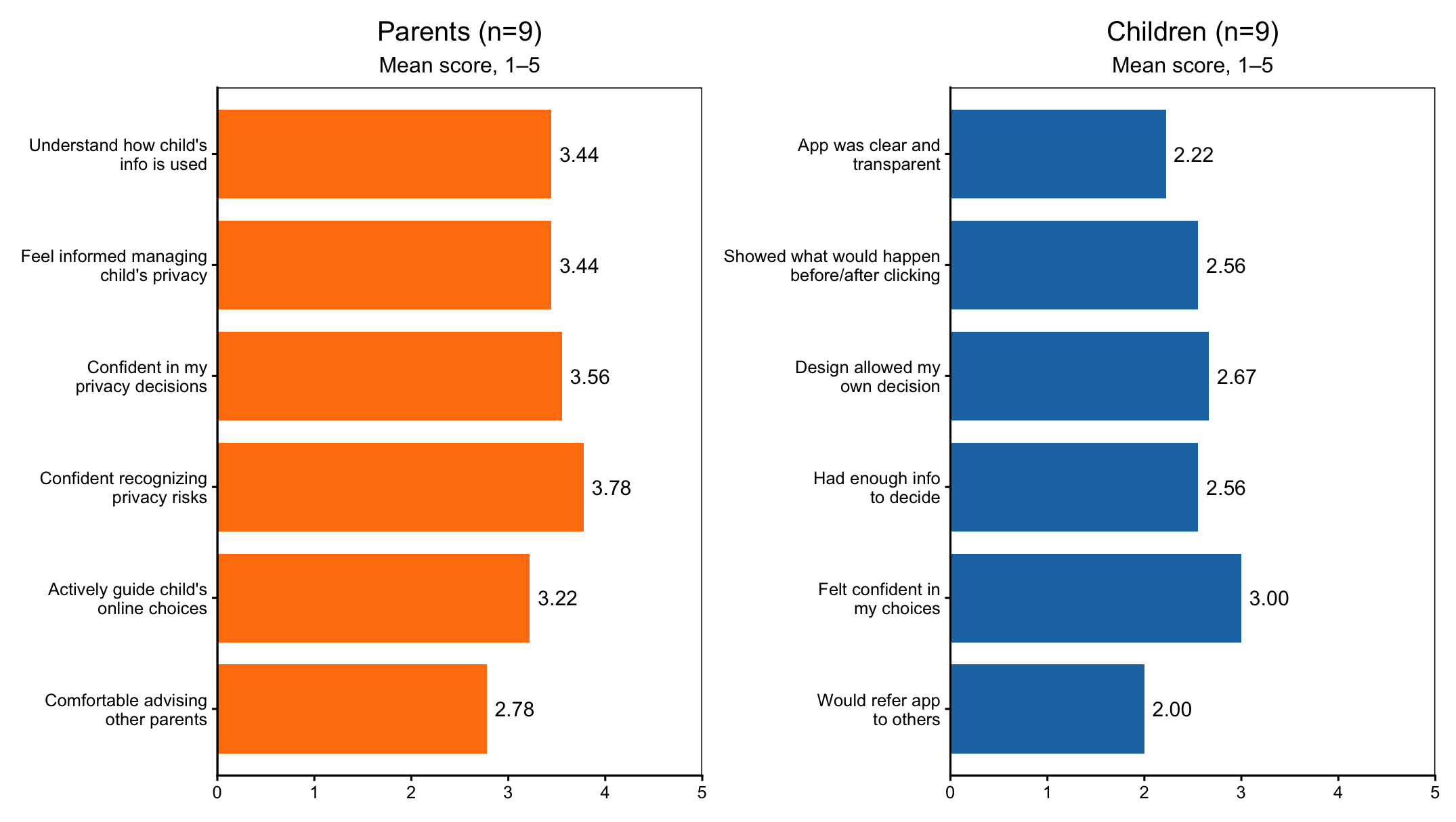}
    \caption{Mean post-survey scores for children and parents across privacy and safety-related measures (1–5 scale)}
    \label{fig:Parents}
\end{figure}

\textbf{Restrictive and limited mediation focused on explicit threats.} In the two pairs with the lowest recognition, parents relied on rules or brief baseline guidance, and their views of privacy did not treat app interfaces as a source of risk. P18 relied on rules rather than conversation: rules about not sharing personal information, and no social media until college. P18 described trading personal information for free services and did not discuss in-app advertising with their children because they did not see it as asking for personal information. P9's parents asked them to avoid unfamiliar things, and P9's responses mirrored this guidance: P9 declined requests they recognized as risky or unwanted, but did not identify the preselected options that shaped those requests. P11's concern about privacy was shaped by identity theft and by data held by devices such as voice assistants, but their mediation with P2 was limited. Their conversations had covered baseline safety, such as not taking pictures with one's clothes off and not keeping secrets, but, as P11 explained (\S\ref{sec:mediation}), they had not yet talked with their children about online things. This guidance addresses an important risk but does not cover location sharing, notification requests, preselected settings, profile visibility, or purchases, the situations in which P2 selected the most prominent option. In both pairs, parents' guidance focused on what to avoid rather than on how interfaces steer choices. These children were also less likely to recognize such design strategies.

\textbf{When Parental Views and Mediation Did Not Explain Recognition.} P16 described their privacy conversations with P7 as limited, yet P7 recognized every deceptive pattern. P16 described P7 as a child who asks questions when purchasing items or when they do not know something, and P7's explanation of data brokers closely echoed P16's own view that websites sell information to third parties: \textit{``they take your data, and then they like they can sell to like, more to, so like scam callers can target you''}. We cannot tell from our data whether P7 learned this from their parent. P7's case suggests that parents' views may reach children through everyday talk or the child's own questions, even when parents describe little privacy education. P13's teaching was incident-driven, arising mainly when P4 encountered something unfamiliar, such as a pop-up advertisement; P4 still recognized the deceptive patterns in five scenarios and partially in the sixth. P8's family drew on workplace security incidents and Cub Scout instruction, yet P8's recognition of fake urgency came from their own experience on Roblox. These pairs suggest that children's own platform experiences and questions, alongside parents' views and mediation, shaped how children recognized deceptive patterns.

\subsection{Post-Survey Results}
\label{sec:survey}
Figure~\ref{fig:Parents} shows mean post-survey scores (5-point scale). Because the child and parent questionnaires asked different questions, we report them separately. Children (n=9), asked about the walkthrough interface, rated their confidence in their own choices highest (M = 3.00) and their willingness to recommend the app to others lowest (M = 2.00). Their ratings of the interface's transparency were also low, from M = 2.22 for whether the app was clear and transparent to M = 2.56 for whether it showed what would happen before or after clicking. Parents (n=9), asked about their own privacy knowledge, were most confident in recognizing privacy risks affecting their child (M = 3.78) and least comfortable advising other parents (M = 2.78).

\section{Discussion}

\subsection{From Privacy Rules to Privacy Reasoning}
In answering RQ1, our results show that children can recognize deceptive patterns, although this ability varied across children, consistent with prior work \cite{sousa2023dark,renaud2024we,schafer2024growing,schafer2025if,arif2026understanding}. Our study extends this work by suggesting that parents' privacy conversations were one of several influences on children's understanding, alongside children's own platform experiences. Even children whose parents described limited conversations about privacy were sometimes able to grasp how these designs worked, suggesting that what conversations addressed may matter as much as how often they occurred. Prior work argues that privacy literacy should support individuals in making well-informed decisions when navigating privacy, security, and online safety \cite{afnan2026we,kumar20225ds}. Kumar et al. \cite{kumar20225ds} similarly argue that children should not be taught isolated rules or tips without explaining what ``privacy'' actually means. This echoes Desimpelaere et al. \cite{desimpelaere2020children}, who found that children often fail to understand hidden or implicit data practices when parents rely primarily on restrictive mediation. We observed a similar pattern in the two pairs with the lowest recognition. P11 and P18 relied on rules or brief baseline guidance focused on what children should not do, such as sharing intimate images or personal information, and did not discuss how apps request information. Their children, P2 and P9, missed the preselected options and visual emphasis that steered their choices.

In contrast, in several pairs where parents described explaining why certain information should or should not be shared, such as P15 explaining the data cycle when P6 downloaded an app, children identified the privacy implications of deceptive patterns more readily. These findings suggest that approaches to children's privacy education may underestimate children's ability to engage with complex privacy issues. Rather than treating privacy as a set of binary rules about what children should or should not do, our findings suggest that children can develop a more nuanced understanding when given opportunities to understand the reasoning behind privacy-related decisions. Systematic reviews have found that children can develop critical-thinking skills through explicit instruction, authentic problem-based learning, and meaningful engagement with subject matter through critical dialogue \cite{asadi2025characteristics, jegstad2025approaches}. These findings suggest that privacy education can move beyond simple, binary rules to foster children's ability to reason through complex privacy decisions.

Our findings also suggest that these conversations were not one-directional. In several families, privacy reasoning developed through exchanges that children helped initiate and shape. P15 kept an open door for their children's questions and talked through information requests together as they arose, and P16 noted that it was easier to guide a child who would \textit{``give a talk''} and \textit{``ask things.''} Children also brought their own experiences into these exchanges: P8 recognized fake urgency from a Roblox offer that never expired, and P7's explanation of how data brokers sell information to scam callers closely echoed P16's account of scam calls following online purchases, although our data cannot show whether P7 learned this from their parent. These accounts align with participatory learning approaches, which position parents and children as engaging with digital media together rather than parents acting solely as gatekeepers~\cite{clark2011parental}, and with research showing that young children can reason and solve problems when given opportunities for dialogue and questioning \cite{o2022critical}. Parents may therefore benefit from explaining the reasoning behind privacy decisions and their consequences, including how apps use defaults, button emphasis, and urgency to steer choices, even when children seem too young for complex privacy issues. For deceptive patterns, children's recognition may develop through such shared, ongoing family practices rather than through parental instruction alone.

\subsection{Balancing Privacy Protection and Children's Agency}

Our second research question asked how parents’ privacy views and mediation practices relate to how their children recognize and respond to deceptive patterns. Central to this relationship is how much agency parents believe children should have online. Parental support for children's autonomy is associated with greater self-regulation and competence \cite{grolnick1989parent}, and among teens, approaches that build self-regulation have been proposed as a middle ground to restrictive parental controls \cite{wisniewski2017parental}. Less is known about how much independence is appropriate at ages 8 to 12. Our study offers one view into this question. Parents in our study answered it differently, and most combined restrictive and active mediation \cite{valkenburg2013developing,rasmussen2013theoretical} with ongoing monitoring. In this study, children whose parents explained how apps use data tended to recognize more patterns than those whose parents mainly set rules about what to avoid (Section~\ref{sec:paired}), though P7’s case shows this was not universal. This fits the view that children’s media experiences are shaped by the parent-child relationship and the wider family context, not by a single mediation style \cite{coyne2017parenting,symons2017qualitative}.

Our findings suggest that restriction and discussion may each play a role. Restriction may reduce children’s exposure to harmful interfaces, but on its own it may not equip them to recognize and respond to manipulative practices \cite{livingstone2017maximizing,rodriguez2018study}. Children in our study were able to reason about the choices they encountered, particularly when privacy was a subject of ongoing discussion at home. This points toward participatory learning \cite{clark2011parental}, in which parents and children engage with digital media together rather than parents acting only as gatekeepers. Under this view, parental involvement becomes a way for children to build their own understanding and decision-making skills. This does not mean granting complete autonomy. Children have been found to develop more digital skills when parents combine high levels of active and restrictive mediation \cite{sciacca2022parental}. For deceptive patterns, parents may keep restrictions as protection while talking through why an interface is pushing a choice and how a child might respond.

Supporting these conversations need not fall on parents alone, particularly because app developers may have little incentive to disclose or address their own deceptive designs. Parental control tools and operating-system protections may help reduce children's exposure to harmful content and deceptive patterns, while regulations requiring greater transparency around such designs could provide broader protections. However, these measures cannot fully replace active parental mediation. Children may still encounter deceptive interfaces beyond the reach of these safeguards, making conversations about why an interface is steering them toward a particular choice important for developing their ability to recognize and question manipulative designs. Ultimately, supporting children's resilience to deceptive patterns may depend on collaboration between parents and children, backed by designers, platforms, and policymakers.

\subsection{Ethical Considerations}

This study was approved by the Institutional Review Board at the authors' institution. Guardians provided written informed consent for themselves and their child, including specific consent for recording, and children gave verbal or written assent. Because assent was obtained in the presence of a consenting parent, we told children directly, in age-appropriate language, that they could skip any question or stop at any time without consequence or loss of compensation. Children were told before the walkthrough that the app was not real, and no real personal or payment information was entered. Afterward, we debriefed each parent and child together, reviewing where the child succeeded or struggled and explaining the deceptive patterns in each scenario. Recordings were stored on institution-managed cloud storage accessible only to the research team, participants were identified by pseudonymous IDs, and quotes were de-identified.

\subsection{Limitations and Future Work}
\label{sec:limt}

This study has several limitations. First, the small sample of nine child-guardian pairs limits the transferability of our findings. However, pairing each child with their own guardian allowed us to examine children's reasoning and guardians' perspectives in depth, which a larger but shallower sample could not provide. As an exploratory qualitative study, it was not designed to establish causal relationships or support population-level generalizations. Second, recruitment through schools, university networks, and social media required in-person, English-language participation. Participants were concentrated in one U.S. region and disproportionately affiliated with academic institutions, which may have produced a sample of families who were already privacy-engaged. Even in this sample, some children missed preselected options and visual steering. Finally, the controlled setting may have heightened children's attentiveness compared with naturalistic app use, as researcher presence, recording, and think-aloud instructions may have encouraged more deliberate scrutiny. Our findings may therefore overestimate children's recognition compared with naturalistic app use. That some children still missed deceptive patterns under these conditions suggests they may be even harder to notice in everyday use. Across scenarios, we observed a potential association between the depth of parental privacy discussions and children's reasoning, but given the small, non-representative sample, this pattern should be interpreted as exploratory rather than causal.

\section{Conclusion}

This study examined how children aged 8–12 recognize and respond to privacy-related deceptive patterns in relation to their parents' privacy views and mediation. Children's recognition varied considerably, from identifying how interfaces steered choices and affected privacy to overlooking manipulative cues. In several families, this variation appeared related to parental privacy conversations. Children whose parents explained how apps collect and use data more often reasoned about deceptive designs, whereas those receiving primarily rules about what to avoid sometimes missed preselected options and visual steering. However, this pattern was not universal, as some children recognized most patterns despite limited parental conversations. Most parents combined restrictive and active mediation, although their approaches to online autonomy varied. These exploratory findings suggest that children's resilience to deceptive patterns may depend not only on interface design and development but also on opportunities to develop privacy reasoning through family conversations. Supporting this requires family-centered resources that move beyond privacy rules, alongside parental controls, platform safeguards, and policies that do not place the burden solely on parents.

\bibliographystyle{ACM-Reference-Format}
\bibliography{CHI2026-child/ref}

\appendix
\appendix

\section{Parent Interview Questions}
\label{appendix:parent}

Let me start by telling you a bit about this project and what we're trying to do. Our research team is studying how school-age children and their guardians perceive privacy-related dark patterns. We're especially interested in exploring similarities and differences in their behaviors and perceptions. As part of the study, we’ll ask your child to complete a few tasks using a social media application that our team developed. These tasks will help us see how children respond to and recognize privacy-related dark patterns. Before we begin that part, we’d like to ask you a few questions about your views on privacy, both for yourself and for your child. To help me keep accurate notes, I’d like to audio-record our session. Please know that any feedback or insights you share may be included anonymously in future research papers or presentations. Is it okay if I record our conversation? Thank you. Do you have any questions before we begin? Great, let’s get started.

\textbf{Background}
\begin{enumerate}
    \item Could you tell me a little about yourself?
    \item What is your occupation, and your partner’s occupation if applicable?
    \item How old is your child?
    \item How does your child perform in school? How involved are you in your child’s education?
\end{enumerate}

\textbf{Personal Privacy Questions}

\begin{enumerate}
    \item How important is privacy to you?
What measures do you take to protect your personal data?

    \item Are there certain types of information you’re especially careful about sharing?
    \item Are there any apps, websites, or technologies you avoid because of privacy concerns, such as data leakage and surveillance?
    \item Have you ever had a negative experience related to a privacy breach or data misuse, e.g., phone number or email getting leaked and being scammed? Could you give me an example?
    \item Do you have any preference on organizations or institutions you most trust when using their services, including data sharing, etc.?
\end{enumerate}

\textbf{Privacy Questions Regarding Their Children}

Do you allow your child to use the internet or social media?

\begin{itemize}
    \item If No
    \begin{itemize}
        \item Why do you choose not to allow your child to use the internet or social media?
        \item Are there specific concerns (e.g., safety, privacy, screen time) that influenced your decision?
        \item How do you plan to prepare or educate your child about using the internet safely in the future?
        \item Do you discuss online privacy and digital safety with your child, even if they aren’t currently using the internet?
        \item What values or habits do you hope to instill before they start using social media or going online independently?
        \item At what age (or under what conditions) would you consider allowing your child to start using the internet or social media? And how would you plan to introduce the technology to your child?
    \end{itemize}
    \item If Yes
    \begin{itemize}
        \item Are you aware of the apps or online platforms your child uses regularly? What are they?
What does your child use these platforms for?

        \item How often do you discuss online privacy and safety with your child?
        \item What steps have you taken to educate your child about online privacy? (e.g., parental controls)
        \item Do you involve other caregivers or family members in educating your child about online privacy?
        \item How do you teach your child to recognize and avoid online scams or suspicious requests for information?
        \item Do you monitor your child’s online activities to ensure their safety in online spaces?
        \item Are there particular privacy risks you worry about most for your child?
Has your child ever shared something online that worried you? How did you handle it?

        \item How do you strike a balance between giving your child autonomy online and keeping them safe?
        \item How confident do you feel about your child’s understanding of online privacy?
    \end{itemize}
\end{itemize}

\section{Child Interview Questions}
\label{appendix:child}
Let me start by telling you a bit about this project and what we are trying to do. Our research team is studying how school-age children like you and their guardians perceive privacy-related dark patterns. We are also exploring similarities and differences in their behaviors and perceptions related to this topic. We will ask you to complete a few tasks on a social media application that my team has developed. This will help us understand how well you recognize privacy-related dark patterns. Before we begin that part, we would like to ask you some questions about your views on privacy issues for yourself. To help me keep accurate notes, I would like to audio record our session. Your feedback and insights may be shared anonymously in future research papers and presentations. Is it okay if I record our conversation? Thank you. Do you have any questions before we begin? Let's get started.

\textbf{Questions Before The Experiments}

\begin{enumerate}
    \item What is your name?
    \item How old are you?
    \item Do you use any types of technology, such as computers or smartphones, or social media?
\end{enumerate}

\begin{itemize}
    \item If No
    \begin{itemize}
        \item Would you like to use the internet one day?
        \item What kinds of dangers do you think can happen when people use the internet?
        \item Are you worried about your information being stolen or used by unauthorized parties?
        \item What are some ways you think you can avoid dangers when going online?
        \item Who would you go to if you have any questions regarding the dangers when using the internet?
    \end{itemize}
    \item If Yes
    \begin{itemize}
        \item What types of technology or devices do you use most often?
        \item Do you use any social media platforms? If yes, what platforms do you normally use?
        \item What kinds of personal information (date of birth, home address, name) do you share online?
        \item How important is your privacy to you?
        \item What steps do you take to protect your personal information in a digital environment?
        \item Have you ever encountered any problems or dangers while using the internet or social media? Could you give me an example?
        \item How do you recognize and avoid potential online dangers or tricks?
        \item Do you talk with others (family, friends) about staying safe online?
        \item What would you do if you got tricked or if your personal information was leaked?
    \end{itemize}
\end{itemize}

\textbf{Questions After The Experiments}

\begin{enumerate}
    \item Have you seen anything on social media that made you think about doing something you normally wouldn’t, or made you feel pressured to do something you didn’t want to do?
If yes, can you tell me some things this social media platform has made you think about doing that seemed a little strange, not right, or annoying?

    \item What made it feel strange or not right to you?
    \item Did you still do any of those things, even if they seemed a little strange or fishy to you?
    
Why did you decide to do it (or not do it)?

    \item Will you do any of those things in the future? Why or why not?
\end{enumerate}

\section{Survey Questions}
\label{appendix:survey}
\textbf{Please rate each statement using the scale below:}

1 = Not at all

2 = A little

3 = Somewhat

4 = A lot

5 = Very much

\textbf{Child Survey Items}

\begin{enumerate}
    \item The app was clear and transparent about what was happening.
    \item The app clearly showed what would happen before and/or after I clicked.
    \item The design allowed me to make my own decision.
    \item I had enough information to make an informed choice.
    \item I felt confident in the choices I made.
    \item I would refer this app to others to use.
\end{enumerate}

\textbf{Parent Survey Items}
\begin{enumerate}
    \item I clearly understand how my child’s personal information is used online.
    \item I feel informed about managing my child’s online privacy.
    \item I feel confident in the decisions I make regarding my child’s online privacy.
    \item I am confident in my ability to recognize potential privacy risks that may affect my child.
    \item I actively guide my child in making safe and responsible online choices.
    \item I feel comfortable advising other parents about their child’s online interactions.
\end{enumerate}

\end{document}